\documentclass{emulateapj} 
\usepackage{amsmath}

\newcommand{\privcomm}[1]{(#1, private communication)}

\begin{document}

\title{Observing IMBH--IMBH Binary Coalescences via Gravitational Radiation}
\shorttitle{Observing IMBH--IMBH Binary Coalescences via Gravitational Radiation}
\submitted{Accepted for publication in \apjl}
\author{John M. Fregeau$^1$, Shane L. Larson$^2$, M. Coleman Miller$^{3,4}$, 
  Richard O'Shaughnessy$^1$, Frederic A. Rasio$^1$}
\shortauthors{Fregeau, et al.}
\affil{$^1$Department of Physics and Astronomy, Northwestern University, Evanston, 
  IL 60208; fregeau@alum.mit.edu, oshaughn@northwestern.edu, rasio@northwestern.edu}
\affil{$^2$Center for Gravitational Wave Physics, The Pennsylvania State University, 
  University Park, PA 16802; shane@gravity.psu.edu}
\affil{$^3$University of Maryland, Department of Astronomy, College Park, MD 20742;
  miller@astro.umd.edu}
\affil{$^4$Goddard Space Flight Center, Greenbelt, MD}

\begin{abstract}
Recent numerical simulations have suggested the possibility of forming double
intermediate mass black holes (IMBHs) via the collisional runaway scenario
in young dense star clusters.  The two IMBHs formed would exchange into
a common binary shortly after their birth, and quickly inspiral and merge.
Since space-borne gravitational wave (GW) observatories such as LISA will be able 
to see the late phases of their inspiral out to several Gpc, and LIGO will be able
to see the merger and ringdown out to similar distances, they represent
potentially significant GW sources.  In this Letter we estimate 
the rate at which LISA and LIGO will see their inspiral and merger in young star 
clusters, and discuss the information that can be extracted from the observations.
We find that LISA will likely see tens of IMBH--IMBH inspirals per year, while 
advanced LIGO could see $\sim 10$ merger and ringdown events per year,
with both rates strongly dependent on the distribution of cluster masses and
densities.
\end{abstract}


\section{Introduction}\label{sec:intro}

Observations suggesting the existence of intermediate-mass black holes (IMBHs)
have mounted in recent years.  Ultra-luminous X-ray sources (ULXs)---point X-ray
sources with inferred luminosities $\gtrsim 10^{39}\,{\rm erg}/{\rm s}$---may
be explained by sub-Eddington accretion onto BHs more massive than the maximum
of $\sim 10M_\sun$ expected from stellar core collapse \citep{2004IJMPD..13....1M}.
Similarly, the cuspy velocity dispersion profiles in the centers of the globular clusters
M15 and G1 may also be explained by the dynamical influence of a central IMBH
\citep{2002AJ....124.3255V,2002AJ....124.3270G,2005ApJ...634.1093G}, although 
this conclusion remains somewhat controversial \citep{2003ApJ...582L..21B}.

The most likely formation scenario for an IMBH is the collapse of a very
massive star (VMS), which was formed early in the lifetime of a young star
cluster via a runaway sequence of physical collisions of massive 
main-sequence stars
\citep{1999A&A...348..117P,2001ApJ...562L..19E,2002ApJ...576..899P,2004ApJ...604..632G}.  
This scenario has been studied in detail for star clusters without primordial
binaries, with recent work showing that runaway 
growth of a VMS to $\sim 10^3\,M_\sun$ occurs generically in clusters with deep core collapse 
times shorter than the $\sim 3\,{\rm Myr}$ main-sequence lifetime of the most
massive stars \citep{2006MNRAS.368..141F}.

Due to the computational cost of simulating the more realistic case of star clusters 
with primordial binaries, it is only recently that such simulations have been performed
\citep{2004Natur.428..724P,2006ApJ...640L..39G}.  The work of \citet{2006ApJ...640L..39G}
was the first to systematically study the influence of primordial binaries on the
runaway growth process.  They showed that stellar collisions during binary
scattering interactions offer an alternate channel for runaway growth, with the main
result that clusters with binary fractions larger than $\approx 10\%$ generically produce 
{\em two} VMSs, provided the cluster is sufficiently dense and/or centrally concentrated
to trigger the runaway earlier than $\sim 3\,{\rm Myr}$ in the absence of primordial binaries.
Observations and recent numerical calculations suggest that star clusters may be
born with large binary fractions \citep[$\gtrsim 30\%$;][]{1992PASP..104..981H,2005MNRAS.358..572I},
implying that {\em all} sufficiently dense and massive star clusters could form
multiple VMSs. 

The VMSs formed will undergo core-collapse supernovae and likely become IMBHs 
on a timescale of $\sim 4\,{\rm Myr}$ after cluster formation 
\citep[the lifetime of a VMS is extended slightly by collisional rejuvenation; see, e.g.][]{2006MNRAS.368..141F}.  
After their separate formation,
the two IMBHs will quickly exchange into a common binary via dynamical
interactions.  The IMBH--IMBH binary (IMBHB) will then shrink via dynamical friction
due to the cluster stars, on a timescale
$\sim t_r \langle m \rangle / M_{\rm IMBH} \lesssim 10\,{\rm Myr}$, where $t_r$ is the
core relaxation time, $\langle m \rangle$ is the local average stellar
mass, and $\langle m \rangle / M_{\rm IMBH} \lesssim 10^{-2}$.
Note that since $t_r$ scales inversely with $\langle m \rangle$
for fixed core velocity dispersion and mass density, the dynamical
friction timescale is independent of $\langle m \rangle$ \citep{1987gady.book.....B}.
The IMBHB will then shrink further via dynamical encounters with cluster stars
\citep{1996NewA....1...35Q,2003ApJ...599.1129Y,2005ApJ...618..426M}, until it merges quickly via 
gravitational radiation, on a timescale 
$\approx 1\,{\rm Myr}\,(\sigma_c/20\,{\rm km}\,{\rm s}^{-1})^3(\rho_c/10^5\,M_\sun\,{\rm pc}^{-3})^{-1}(M_{\rm IMBH}/10^3\,M_\sun)^{-1}$,
where $\sigma_c$ is the cluster core velocity dispersion and $\rho_c$ is the core mass density
\citep[][eqs.~{[29]} and {[30]}]{1996NewA....1...35Q}.
This timescale has also been confirmed by numerical scattering calculations 
\privcomm{G\"ultekin}.

\begin{figure}
  \begin{center}
    \includegraphics[width=\columnwidth]{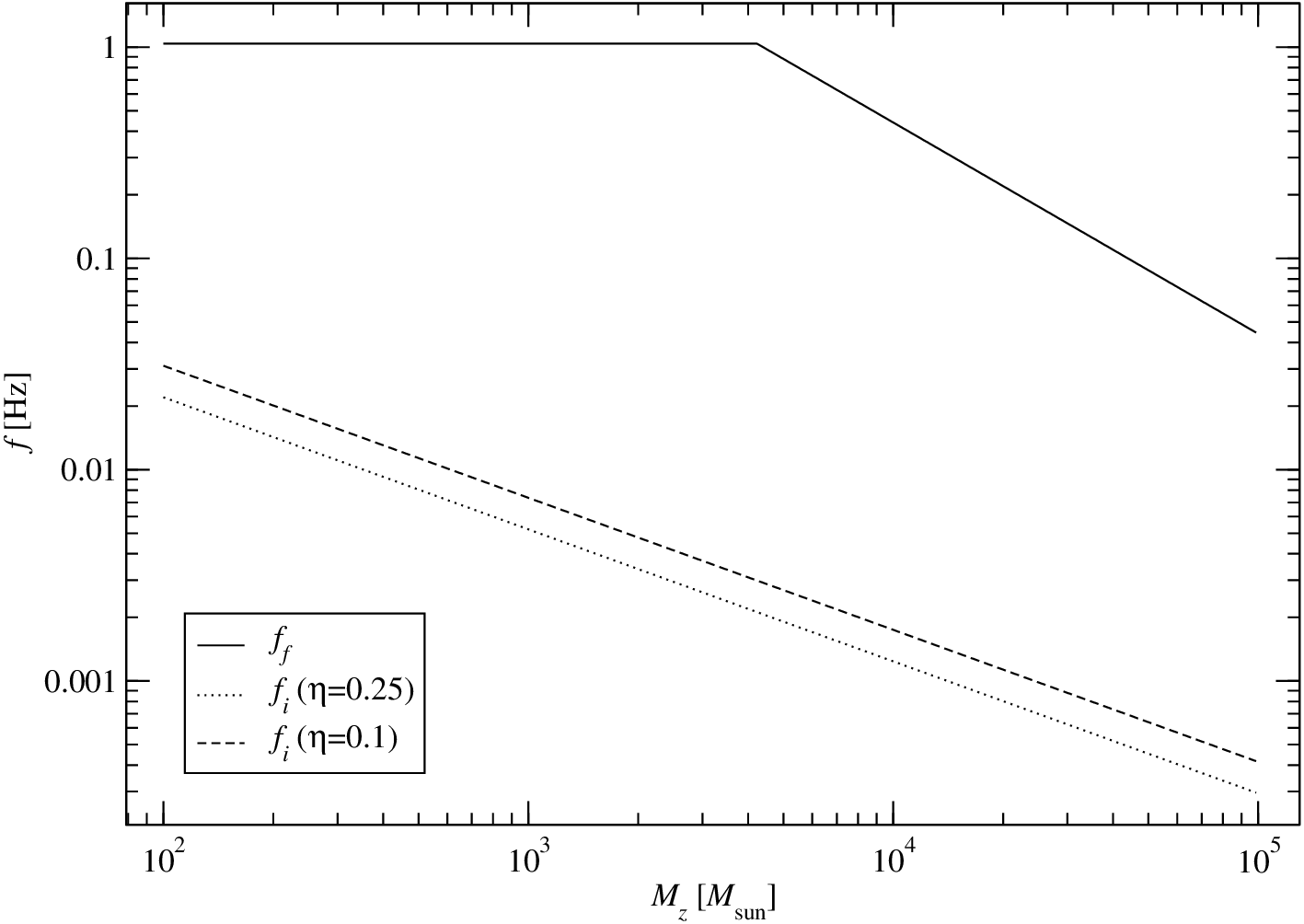}
    \caption{The final GW frequency $f_f$ (see text), and the frequency 
      1 yr prior, $f_i$, for an IMBHB with
      total mass $M$ and reduced mass parameter $\eta$,
      as a function of redshifted binary mass $M_z$, for
      $\eta=0.25$ (equal-mass binary) and $\eta=0.1$ (mass ratio $0.13$).
      (The final frequency is roughly independent of $\eta$.)
      \label{fig:fs}}
  \end{center}
\end{figure}

Only the more massive IMBHBs merge in the LISA band
of $10^{-4}$--$1\,{\rm Hz}$ (redshifted binary mass 
$M_z \equiv (1+z)M \gtrsim 4 \times 10^3 \, M_\sun$, where $M$ is the 
total binary mass).  Fig.~\ref{fig:fs} shows the final
gravitational wave (GW) frequency $f_f$ (the frequency at
the inner-most stable circular orbit if within the LISA frequency range (large $M_z$), otherwise
the maximum LISA frequency of $\approx 1\,{\rm Hz}$ (small $M_z$), 
as in \citet{2004ApJ...611.1080W}),
and the frequency 1 yr prior, $f_i$, as a function of redshifted
mass $M_z$, for the reduced mass parameters $\eta=0.25$ (equal-mass binary)
and $\eta=0.1$ (mass ratio $0.13$) \citep[see, e.g.,][]{2004ApJ...611.1080W}.
For a wide range in $M_z$, the late stages of inspiral clearly
span the LISA ``sweet spot'' (roughly a decade centered on $10^{-2.2}\,{\rm Hz}$),
implying that LISA could easily detect the chirp signal, enabling 
a measurement of the masses of the binary members.  Such an observation would 
be direct evidence for an IMBH.

In Sec.~\ref{sec:lisa} we estimate the rate at which LISA will observe inspiral of
IMBHBs in young star clusters.  In Sec.~\ref{sec:ligo} we estimate the rate at which 
LIGO will observe their merger and ringdown.  Finally, in Sec.~\ref{sec:discussion} 
we discuss the observational consequences.

\begin{figure}
  \begin{center}
    \includegraphics[width=\columnwidth]{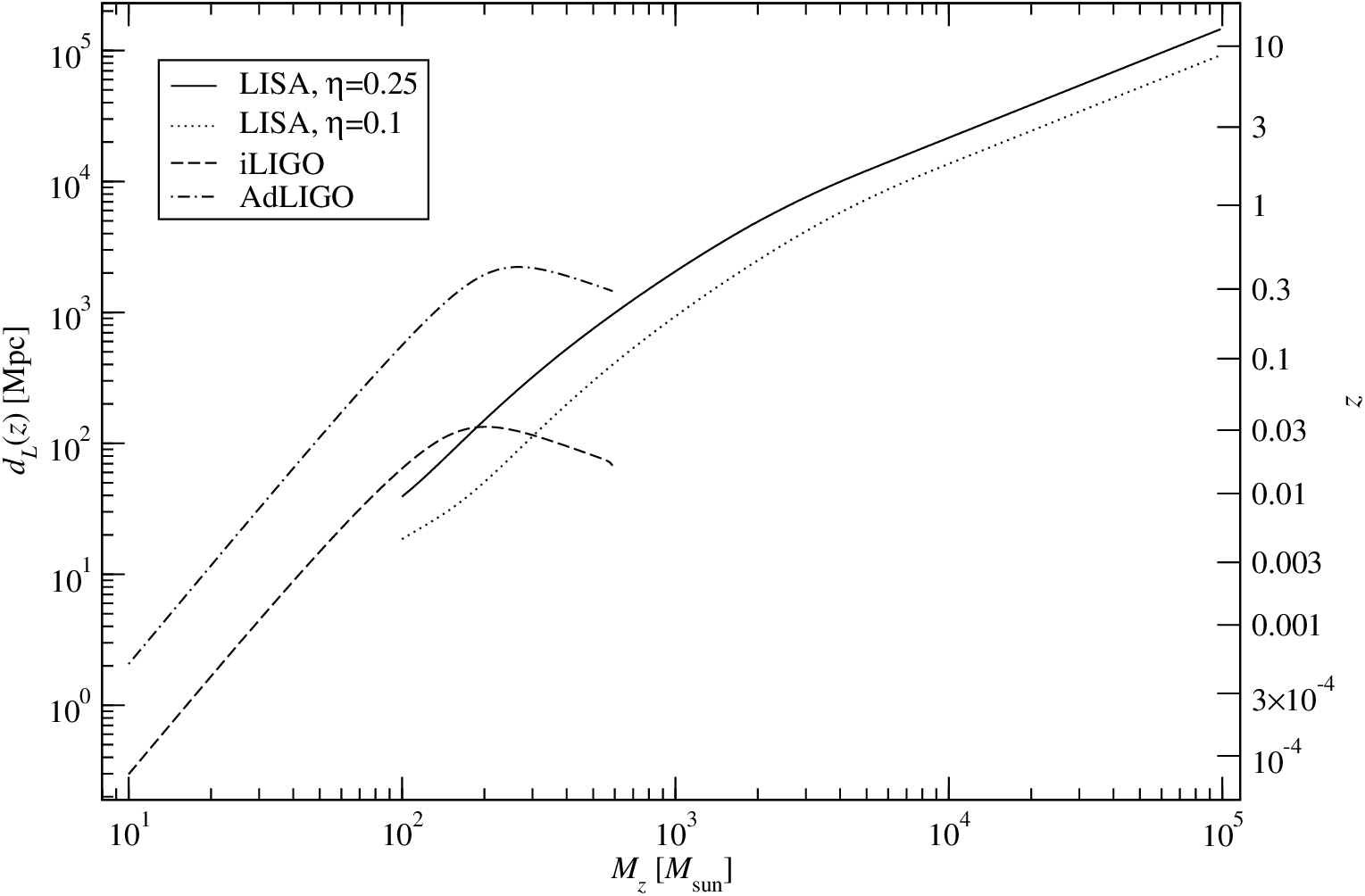}
    \caption{Luminosity distance, $d_L(z)$, to which an IMBHB of
      total mass $M$ and reduced mass parameter $\eta$ can be seen
      via its inspiral with LISA with $S/N=10$ for a 1 yr integration,
      and via its merger and ringdown
      with $S/N=8$ for iLIGO and AdLIGO,  as a function
      of the redshifted mass $M_z$.  The corresponding redshift
      (calculated using the WMAP year 3 cosmological parameters,
      as discussed in the text) is shown on the right vertical axis.
      \label{fig:dlvsm}}
  \end{center}
\end{figure}

\section{Estimating the LISA detection rate}\label{sec:lisa}

We first need to know the distance to which LISA can see IMBHB inspirals.  
Following the techniques in \citet{2004ApJ...611.1080W} and 
\citet{1998PhRvD..57.4535F}, we adopt the latest LISA sensitivity curve \citep{SCG},
including confusion noise from Galactic white dwarf binaries \citep{benderhils1997},
and calculate the maximum luminosity distance, $d_L(z)$, to which an IMBHB
of total mass $M$ and reduced mass parameter $\eta$ can be
seen with $S/N=10$ for a 1 yr integration.
The results are shown in Fig.~\ref{fig:dlvsm} as a function of $M_z$, for $\eta=0.25$ and $\eta=0.1$.
Note that the results of \citet{2006ApJ...640L..39G} show that the masses of the 
IMBHs never differ by more than a factor of a few ($\eta \gtrsim 0.15$).  
Thus LISA will be able to see typical IMBHBs ($M \sim 10^3\,M_\sun$) out to a few Gpc.

With this information in hand, we first make a crude
estimate of the LISA event rate.  Following \citet{2002ApJ...581..438M}, 
we write for the total rate
\begin{equation}\label{eq:a}
  R \equiv
  \frac{dN_{\rm event}}{dt} = 
  \left(\int_0^{z_{\rm max}} \frac{dV_c}{dz} dz \right)
  \frac{dN_{\rm cl}}{dV}
  g
  \frac{1}{t_U} \, .
\end{equation}
The first factor, $\int_0^{z_{\rm max}} (dV_c/dz) dz$, is the integrated comoving
volume of space in which LISA is sensitive to the events.  The second factor, $dN_{\rm cl}/dV$,
is the number density of star clusters sufficiently massive
to form IMBHBs.  Since the {\em globular} clusters we currently
see were likely at least a few times more massive at formation \citep{2001ApJ...550..691J},
we set this factor to the current density of globular clusters in the local
universe, $dN_{\rm cl}/dV \approx 8 h^3\,{\rm Mpc}^{-3}$ \citep{2000ApJ...528L..17P}.  
The third factor, $g$, is the fraction of sufficiently massive clusters that have a large
enough binary fraction and initial central density to produce IMBHBs.  
Since initial cluster structural parameters are largely unknown, we treat $g$ as a parameter.
The fourth factor is the event rate per IMBHB-producing cluster, taken to be 
one divided by the age of the universe, since only one IMBHB
is formed per cluster over its lifetime.  We adopt a $\Lambda$CDM cosmology,
with parameters $\Omega_M=0.24$, $\Omega_\Lambda=0.76$, and $h=0.73$, for
which $t_U=13.8\,{\rm Gyr}$ \citep{astro-ph/0603449}.  Putting this together for 
$d_L=4.9\,{\rm Gpc}$ ($z_{\rm max}=0.79$), the distance to which LISA can see
IMBHBs with $M=2\times 10^3\,M_\sun$, eq.~(\ref{eq:a}) gives 
$R \approx 1(g/0.1)\, {\rm yr}^{-1}$.

Writing down a generalized form of the rate integral in eq.~(\ref{eq:a}) is 
straight-forward.  Since the time between cluster formation and IMBHB merger 
is $\ll t_U$, we assume that the merger is coincident with cluster formation.  
Thus the rate integral is
\begin{multline}\label{eq:b}
  R \equiv
  \frac{dN_{\rm event}}{dt_o} = 
  \int_0^{z_{\rm max}}
  \frac{d^2M_{\rm SF}}{dV_c dt_e}
  g_{\rm cl}
  g\\
  \times \frac{dt_e}{dt_o} 
  \frac{dV_c}{dz}
  \int_{M_{\rm cl,min}(z)}^{M_{\rm cl,max}}
  \frac{d^2N_{\rm cl}}{dM_{\rm SF,cl} dM_{\rm cl}}
  \, dM_{\rm cl} \, dz \, .
\end{multline}
Here $R \equiv dN_{\rm event}/dt_o$ is the event rate observed at $z=0$
by LISA, $d^2M_{\rm SF}/dV_c dt_e$ is the star formation rate (SFR) in 
mass per unit of comoving volume per unit of local time, 
$g_{\rm cl}$ is the fraction of star forming mass that goes
into star clusters more massive than $10^{3.5}\,M_\sun$ 
(generally a function of $z$), $g$ is as above, and
$d^2N_{\rm cl}/dM_{\rm SF,cl} dM_{\rm cl}$ is the distribution function 
of clusters over individual cluster mass $M_{\rm cl}$ and
total star forming mass in clusters $M_{\rm SF,cl}$.  Finally, $dt_e/dt_o$ is simply
$(1+z)^{-1}$, and $dV_c/dz$ is the rate of change of comoving volume
with redshift, which is a function of cosmological parameters \citep{astro-ph/9905116}.
Note that we set $z_{\rm max}=5$, since this is roughly the limit to which the
cosmic SFR can be traced.  Thus the integral in eq.~(\ref{eq:b})
should be considered a mild lower limit to the true rate.  We now discuss each 
element in eq.~(\ref{eq:b}) in more detail.

Following \citet{2001ApJ...548..522P}, we adopt three different choices
for the SFR:
\begin{equation}\label{eq:sfr}
  \left(\frac{d^2M}{dV_c dt}\right)_{{\rm SF}i} = C_i h_{65} F(z) G_i(z) M_\sun\,{\rm yr}^{-1}\,{\rm Mpc}^{-3} \, ,
\end{equation}
with $i=1,2,3$ denoting the different rates, 
$C_i$ a constant, $G_i(z)$ a function of $z$, $h_{\rm 65}=h/0.65$, and 
$F(z)=[\Omega_M(1+z)^3+\Omega_k(1+z)^2+\Omega_\Lambda]^{1/2}/(1+z)^{3/2}$.
The first is from \citet{2000MNRAS.312L...9M}, with 
$C_1=0.3$ and $G_1(z)=e^{3.4z}/(e^{3.8z}+45)$, which peaks between $z=1$ and 2 
and decreases at larger redshift.  The second is from \citet{1999ApJ...519....1S}, with
$C_2=0.15$ and $G_2(z)=e^{3.4z}/(e^{3.4z}+22)$, which is roughly constant for 
$z \gtrsim 2$.  The third is from \citet{1999ApJ...512L..87B}, with
$C_3=0.2$ and $G_3(z)=e^{3.05z-0.4}/(e^{2.93z}+15)$, 
which increases above $z \approx 2$.

Measuring the fraction of star-forming mass in clusters is 
difficult for anywhere but the local universe.
Similarly, while we know reasonably well the initial cluster conditions required 
to form an IMBHB \citep{2006ApJ...640L..39G},
we know much less well the distribution of cluster properties at birth.
We therefore treat $g_{\rm cl}$ and $g$ as parameters, taking $g_{\rm cl} = 0.1$ and
$g=0.1$ somewhat arbitrarily as canonical values.

Assuming that the spectrum of cluster masses is neither a function of cosmic
epoch nor the total star forming mass available for clusters, the factor
$d^2N_{\rm cl}/dM_{\rm SF,cl} dM_{\rm cl}$ can be separated as
\begin{equation}
  \frac{d^2N_{\rm cl}}{dM_{\rm SF,cl} dM_{\rm cl}} = 
  \frac{f(M_{\rm cl})}{\int M_{\rm cl} f(M_{\rm cl}) dM_{\rm cl}} \, ,
\end{equation}
where $f(M_{\rm cl})$ is the (normalized) distribution function of cluster masses.  For this
we adopt the power-law form observed for young star clusters in the Antennae, which is 
thought to be universal: $f(M_{\rm cl}) \propto M_{\rm cl}^{-2}$ \citep{1999ApJ...527L..81Z},
with a lower limit of $10^{3.5}\,M_\sun$ and an upper limit of $10^7\,M_\sun$.

It is the limit $M_{\rm cl,min}(z)$ in eq.~(\ref{eq:b}) that 
encodes all information about the detectability of an IMBHB inspiral by 
LISA.  Specifically, the redshift to which LISA can see the inspiral is a function
of the binary mass, which is itself a function of the host cluster mass.  Adopting
an efficiency factor $f_{\rm GC}$ for the fraction of
cluster mass going into the IMBHB, this
relationship is inverted to obtain $M_{\rm cl,min}(z)$.  
Recent numerical work shows that the efficiency factor is 
$f_{\rm GC}\approx 2\times 10^{-3}$, independent of cluster initial 
conditions \citep{2004ApJ...604..632G}, which we take as our canonical value.  
At low redshift, $M_{\rm cl,min}(z)$ is clamped at the value $M_{\rm cl}=200\,M_\sun/f_{\rm GC}$, 
set by adopting the definition that an IMBH have mass $\geq 10^2\,M_\sun$.
At high redshift ($z > 5$, so not relevant to our calculation), 
$M_{\rm cl,min}(z)$ is clamped at the value of $10^7\,M_\sun$ from the cluster
mass function; in other words no cluster is sufficiently massive to produce an
IMBHB massive enough to be observable by LISA, so the integral is zero.

\begin{figure}
  \begin{center}
    \includegraphics[width=\columnwidth]{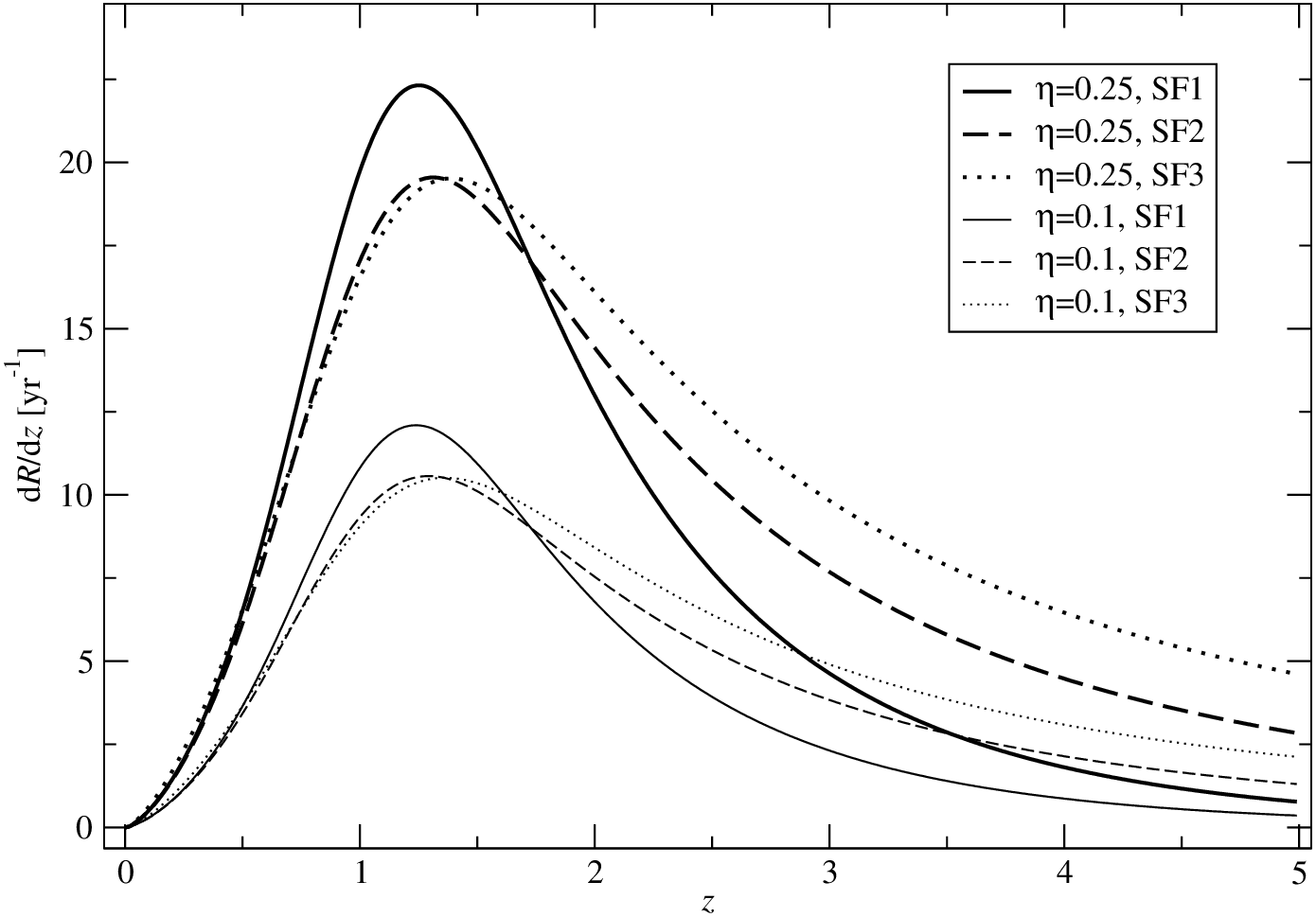}
    \caption{Integrand of the rate integral in eq.~(\ref{eq:b}) for
      the three different SFRs in eqs.~(\ref{eq:sfr}), 
      for $\eta=0.25$ and $\eta=0.1$.
      \label{fig:dRdz}}
  \end{center}
\end{figure}

We numerically integrated eq.~(\ref{eq:b}) for the different SFRs in eqs.~(\ref{eq:sfr}),
for $S/N=10$ and an integration time of 1 yr, to find that the rate is 
\begin{equation}\label{eq:rate1}
  R(\eta=0.25)\approx 40\mbox{--}50 
  \left(\frac{g_{\rm cl}}{0.1}\right) \left(\frac{g}{0.1}\right) 
       {\rm yr}^{-1} \, ,
\end{equation}
with the spread in the coefficient from the different
SFRs.  The coefficient decreases to $20\mbox{--}25$ for 
$\eta=0.1$.  The rate is dominated by clusters in the mass range 
$10^6$--$10^{6.5}\,M_\sun$ (IMBHB mass $2 \times 10^3$--$6\times 10^3\,M_\sun$), 
with more than half the contribution to the rate coming 
from this mass range, for both $\eta=0.1$ and $\eta=0.25$, and for all 
three SFRs in eq.~(\ref{eq:sfr}) (except SF3 for $\eta=0.1$).
Note that eq.~(\ref{eq:b}) is only strictly valid when the source is 
visible by the instrument for less than the integration
time.  This turns out not to be precisely correct.  A typical IMBHB
with mass $M=f_{\rm GC} 10^{6.25}\,M_\sun$ takes roughly 4 years to cross
the LISA band from the edge of the white dwarf confusion knee at $\approx 2\,{\rm mHz}$ 
to the upper edge of the band at $\approx 1\,{\rm Hz}$.  Thus the rate presented in
eq.~(\ref{eq:rate1}) is an underestimate by of order a factor of a few.

Fig.~\ref{fig:dRdz} shows
the integrand of the rate integral in eq.~(\ref{eq:b}) for
the three different SFRs in eqs.~(\ref{eq:sfr}), 
for $\eta=0.25$ and $\eta=0.1$.  Most events originate
from $z \sim 1$.  Unfortunately, neither $R$ nor $dR/dz$ is particularly
sensitive to the cosmic SFR, with $dR/dz$ decreasing
quickly above $z \approx 2$ even when the SFR is increasing
(as in SF3).  Thus observations of IMBHB inspirals will not be very
informative about the cosmic SFR.
However, they will likely yield a handle on the fraction of star formation 
that is in compact massive clusters.

\section{Estimating the LIGO detection rate}\label{sec:ligo}

Shortly after the two IMBHs merge, the merger product can be well
described as a single perturbed black hole, emitting GWs at its quasinormal frequencies.
Largely falling within the initial and advanced LIGO (iLIGO and AdLIGO) sensitivity
bands, the merger and ringdown waves will likely carry a few percent of the rest mass
energy of the hole \citep[see, e.g.][]{1998PhRvD..57.4535F}.  
Numerical simulations suggest that a merging pair of nonspinning equal-mass
black holes will emit a fraction $\epsilon \simeq 0.03$ of their rest mass in merger and
ringdown GWs, forming a black hole with spin parameter $a \simeq 0.7$ 
\citep{Lazarus,NR:Brownsville,gr-qc/0602026}.  Under these conditions, the
ringdown frequency is given by \citep[see {eq.~[3.17]} of][]{1998PhRvD..57.4535F}
\begin{equation}
  f \approx \frac{c^3}{2 \pi G M_z} (1-0.63(1-a)^{3/10})
  \approx 180 \left(\frac{M_z}{10^{2}M_\odot}\right)^{-1} {\rm Hz} \, .
\end{equation}
We can express the distance to which we are
sensitive to ringdown waves at signal-to-noise ratio $\rho$ as
\begin{equation}
d_L(z) = \left(\frac{2 \epsilon M_z}{5 \pi^2\rho^2 f^2 S(f)}\right)^{1/2} \, ,
\end{equation}
where $S(f)$ is the spectral noise density of LIGO.
Combining this expression with the concordance cosmological model and
iLIGO and AdLIGO sensitivity curves, we find the range to
which LIGO can detect ringdown shown in Fig.~\ref{fig:dlvsm}.

To obtain a conservative estimate for the rate at which iLIGO and AdLIGO
could detect these mergers with a ringdown-only search,
we use eq.~(\ref{eq:a}) with a
moderately optimistic range of $d_L \approx 100\,{\rm Mpc}$ for iLIGO and 
$d_L = 2\,{\rm Gpc}$ for AdLIGO.  The expected detection rate is then
$10^{-4} (g/0.1)\,{\rm yr}^{-1}$ and $1 (g/0.1)\,{\rm yr}^{-1}$, 
for iLIGO and AdLIGO, respectively.  More detailed estimates using 
machinery analogous to eq.~(\ref{eq:b}) increase these estimates by roughly 
an order of magnitude, making the rate for AdLIGO 
$10 (g_{\rm cl}/0.1) (g/0.1)\,{\rm yr}^{-1}$.



\section{Discussion}\label{sec:discussion}

It appears likely that LISA will see tens of IMBHB inspiral events per year,
while AdLIGO could see $\sim 10$ merger and ringdown events per year,
with both rates strongly dependent on the distribution of cluster masses and
densities.  Detection of an IMBHB would have profound implications.  A match-filtered
observation of the inspiral would yield the redshifted masses of the black
holes, directly confirming the existence of IMBHs.  It would also yield the
luminosity distance to the source; with enough observations, constraints could
be placed on the cosmic history of star formation in dense, massive clusters.
Detection of the ringdown signal from the merger product will additionally 
yield its spin, which may provide insight into its formation history.

Typical IMBHBs spend $\gtrsim 10^6\,{\rm yr}$ inspiraling
through the LISA band, with nearly all of that time spent at low frequencies 
($\lesssim 10^{-3}\,{\rm Hz}$).  In the low frequency
region they will thus appear as a large number of monochromatic
sources, possibly contributing to confusion noise and increasing the noise 
floor \citep[e.g.,][]{2003MNRAS.346.1197F}.  A detailed calculation of this
is beyond the scope of this Letter.  However,
we note that if their contribution is similar in magnitude to that of Galactic
compact object binaries \citep{benderhils1997}, the rates predicted in
eq.~(\ref{eq:rate1}) would decrease by about 20\%.

\acknowledgments

We thank Kayhan G\"ultekin for providing the numerically-calculated timescale for an IMBHB to shrink
by stellar dynamical encounters after formation.
We also thank the anonymous referee for many helpful comments on the manuscript.
JMF and FAR acknowledge support from NASA grant NNG04G176G and NSF grant PHY-0245028.
SLL acknowledges support from NASA award NNG05GF71G.
MCM was supported in part by NASA grant NAG 5-13229 and by the Research Associateships
Programs Office of the National Research Council and Oak Ridge Associated Universities.
ROS acknowledges support from NSF grants PHYS-0121416 and PHYS-0353111.

\bibliographystyle{apj}
\bibliography{apj-jour,main}

\end{document}